\documentclass[sigconf]{acmart}

\usepackage{booktabs} 
\usepackage{amsmath}
\usepackage{amssymb}
\usepackage[ruled, linesnumbered]{algorithm2e}
\usepackage{amsmath}
\usepackage{xcolor}
\usepackage{multirow}
\usepackage{graphicx}
\usepackage{subfigure}
\usepackage{algorithmic}
\usepackage{amsfonts}

\setcopyright{rightsretained}

\begin{document}
\title{A Neural Influence Diffusion Model for Social Recommendation}


\author{Le Wu}
\affiliation{%
  \institution{Hefei University of Technology}
}
\email{lewu@hfut.edu.cn}

\author{Peijie Sun}
\affiliation{%
  \institution{Hefei University of Technology}
}
\email{sun.hfut@gmail.com}

\author{Yanjie Fu}
\affiliation{%
  \institution{Missouri University of Science and Technology}
}
\email{fuyan@mst.edu}

\author{Richang Hong}
\affiliation{%
  \institution{Hefei University of Technology}
}
\email{hongrc.hfut@gmail.com}

\author{Xiting Wang}
\affiliation{%
  \institution{Microsoft Research}
}
\email{xitwan@microsoft.com}

\author{Meng Wang}
\affiliation{%
  \institution{Hefei University of Technology}
}
\email{eric.mengwang@gmail.com}

\begin{abstract}

Precise user and item embedding learning is the key to building a successful recommender system. Traditionally, Collaborative Filtering~(CF) provides a way to learn user and item embeddings from the user-item interaction history. However, the performance is limited due to the sparseness of user behavior data. With the emergence of online social networks, social recommender systems have been proposed to utilize  each user's local neighbors' preferences to alleviate the data sparsity for better user embedding modeling. We argue that, for each user of a social platform, her potential embedding is influenced by her trusted users, with these trusted users are influenced by the trusted users' social connections. As social influence recursively propagates and diffuses in the social network, each user's interests change in the recursive process. Nevertheless, the current social recommendation models simply  developed static models by leveraging the local neighbors of each user  without simulating the recursive diffusion in the global social network, leading to suboptimal recommendation performance. In this paper, we propose a deep influence propagation model to stimulate how users are influenced by the recursive social diffusion process for social recommendation.  For each user, the diffusion process starts with an initial embedding that fuses the related features and a free user latent vector that captures the latent behavior preference. The key idea of our proposed model is that we design a layer-wise influence propagation structure to model how users' latent embeddings evolve as the social diffusion process continues. We further show that our proposed model is general and could be applied when the user~(item) attributes or the social network structure is not available.  Finally, extensive experimental results on two real-world datasets clearly show the effectiveness of our proposed model\footnote{code: https://github.com/PeiJieSun/diffnet}, with more than 13\%  performance improvements over the best baselines for top-10 recommendation on the two datasets.

\end{abstract}

\maketitle

\section{Introduction}

By providing personalized item suggestions for each user, recommender systems have become a cornerstone of the E-commerce shopping experience ~\cite{adomavicius2005toward,su2009survey}.  Among all recommendation algorithms, learning low dimensional user and item embeddigs is a key building block that have been widely studied~\cite{NIPS2008probabilistic,koren2009matrix,rendle2012factorization}. With the learned user and item embeddings, it is convenient to approximate the likelihood  or the predicted preference that a user would give to an item by way of a simple inner product between the corresponding user embedding and item embedding.

Many efforts have been devoted to designing sophisticated models to learn precise user and item embeddings.
In the typical collaborative filtering scenario with user-item interaction behavior, the latent factor based approaches have received great success in both academia and industry due to their relatively high performance~\cite{koren2009matrix,UAI2009bpr,NIPS2008probabilistic,rendle2012factorization}. 
Though successful, the recommendation performance is unsatisfactory due to the sparseness of user-item interaction data. As sometimes users and items are associated with features, factorization machines generalize most latent factor models with an additional linear regression function of user and item features~\cite{rendle2012factorization}. Recently, researchers also
designed more advanced neural models based on latent factor models and FMs~\cite{WWW2017neural,guo2017deepfm}. E.g., NeuMF extends over latent factor based models by modeling the complex relationships between user and item embedding with a neural architecture~\cite{WWW2017neural}. These deep embedding models advance the performance of previous shallow embedding models. Nevertheless, the recommendation performance is still hampered by the sparse data.

Luckily, with the prevalence of online social networks, more and more people like to express their opinions of items in these social platforms.  The social recommender systems have emerged as a promising direction, which leverage the social network among users to alleviate the data sparsity issue  and enhance recommendation performance~\cite{WSDM2011recommender,TKDE2014scalable,guo2015trustsvd,tang2013social}.  These social recommendation approaches are based on the social influence theory that states connected people would influence each other, leading to the similar interests among social connections~\cite{Nature2012,PNAS2012social,KDD2008influence}. E.g., social regularization has been empirically proven effective for social recommendation, with the assumption that connected users would share similar latent embeddings~\cite{Recsys2010matrix,WSDM2011recommender,TKDE2014scalable}. TrustSVD++ extended the classic latent factor based models by incorporating each user's trusted friends' feedbacks to items as the auxiliary feedback of the active user~\cite{guo2015trustsvd}.  All these works showed performance improvement by considering the first-order local neighbors of each user. Nevertheless, we argue that, for each user, instead of the interest diffusion from a user's neighbors to this user at one time, the social diffusion presents a dynamic recursive effect to influence each user's embedding. In detail, as the social influence propagation process begins~(i.e., the diffusion iteration $k=1$), each user's first latent embedding is influenced by the initial embeddings of her trusted connections. With the recursive influence diffuses over time, each user's latent embedding at $k$-th iteration is influenced by her trusted neighbors at the $(k-1)$-th iteration. Therefore, the social influence recursively propagates and diffuses in the social network. Correspondingly, each user's interests change in the recursive process. Precise simulating the recursive diffusion process in the global social network would better model each user's embedding, thus improve the social recommendation performance.

In this paper, we propose DiffNet: an Influence \emph{Diff}usion neural network based  model to stimulate the recursive social influence propagation process for better user and item  embedding modeling in social recommendation. The key idea behind the proposed model is a carefully designed layer-wise influence diffusion structure for users, which models how users' latent embeddings evolve as the social diffusion process continues. Specifically, the diffusion process starts with an initial embedding for each user on top of the fusion of each user's features and a a free user latent vector that captures the latent behavior preference. For the item side, as items do not propagate in the social network, each item's embedding is also fused by the free item latent embedding and the item features. With the influence diffuses to a predefined $K$-th diffusion step, the $K$-th layer user interest embedding is obtained. In fact, with the learned user and item embedddings, DiffNet can be seamlessly incorporated into classical CF models, such as BPR and SVD++, and efficiently trained using SGD.

We summarize the contributions of this paper as follows:

\begin{itemize}
  \item We propose a DiffNet model  with a layer-wise influence propagation structure to model the recursive dynamic social diffusion in social recommendation. Besides, DiffNet has a  fusion layer such that each user and each item could be represented as an embedding that encompasses both the collaborative and the feature content information.
  \item We show that the proposed DiffNet model is time and storage efficient in comparison to most embedding based recommendation models. The proposed model is a generalization of many related recommendation models and it is flexible  when user and item attributes are not available.
  \item Experimental results on two real-world datasets clearly show the effectiveness of our proposed DIP model.  DiffNet outperforms more than \textbf{13.5}\%  on \textit{Yelp} and  \textbf{15.5}\% on \textit{Flickr} for top-10 recommendation compared to the the baselines with the best performance.
\end{itemize}

\section{Problem Definition and Preliminaries}

\subsection{Problem Definition}
In a social recommender system, there are two sets of entities: a user set {\small$U$~($|U|\!=\!M$)}, and an item set {\small$V$~($|V|\!=\!N$)}. Users interact with items to show their preference. As the implicit feedback~(e.g., watching an movie, purchasing an item, listening to a song ) are more common in, we also consider the recommendation scenario with implicit feedback~\cite{UAI2009bpr}. Let {\small $\mathbf{R}\in\mathbb{R}^{M\times N}$} denote users' implicit feedback based rating matrix, with $r_{ai}\!=\!1$ if user $a$ is interested in item $i$, otherwise it equals 0. The social network can be represented as a user-user directed graph {\small$\mathcal{G}=[U, \mathbf{S}\in\mathbb{R}^{M\times M}]$}, with $U$ is the user set and {\small $\mathbf{S}$} represents the social connections between users. If user $a$ trusts or follows user $b$, $s_{ba}=1$, otherwise it equals 0. If the social network is undirected, then user $a$ connects to user $b$ denotes $a$ follows $b$, and $b$ also follows $a$, i.e., $s_{ab}=1 \wedge s_{ba}=1$. Then, each user $a$'s  ego social network, i.e.,  is the i-th column~($\mathbf{s}_a$) of {\small $\mathbf{S}$}.  For notational convenience, we use
$S_a$ to denote the userset that $a$ trusts, i.e., $S_a=[b| s_{ba}=1]$.

Besides, each user $a$ is associated with real-valued attributes~(e.g, user profile), denoted as $\mathbf{x}_a$ in the user attribute matrix {\small $\mathbf{X}\in\mathbb{R}^{d1\times M}$}. Also, each item $i$ has an attribute vector $\mathbf{y}_i$~(e.g., item text representation, item visual representation) in item attribute matrix {\small $\mathbf{Y}\in\mathbb{R}^{d2\times N}$}. Without confusion, we use $a, b, c$ to denote users and $i, j, k$ to denote items.  The matrices are denoted with capital bold letters, and vectors with small bold letters. Then, the social recommendation problem can be defined as:

\begin{definition}  [{\small \textbf{SOCIAL RECOMMENDATION}}] Given a rating matrix {\small$\mathbf{R}$}, a social network {\small$\mathbf{S}$}, and associated real-valued feature matrix {\small $\mathbf{X}$} and {\small $\mathbf{Y}$} of users and items, our goal is to predict users' unknown preferences to items as:
$\hat{R}=f(\mathbf{R},\mathbf{S}, \mathbf{X}, \mathbf{Y})$, where {\small$\hat{R}\in\mathbb{R}^{M\times N}$} denotes the predicted preferences of users to items.
\end{definition}

\subsection{Preliminaries}

\textbf{Classical Embedding Models.} Given the user-item rating matrix {\small$\mathbf{R}$}, the latent embedding based models are among the most successful approaches to capture the collaborative information for building the recommender systems~\cite{UAI2009bpr,koren2009matrix,rendle2012factorization}.  Specifically, these latent embedding models embed both users and items in a low latent space, such that each user's predicted preference to an unknown item turns to the inner product between the corresponding user and item embeddings as:

\vspace{-0.2cm}
 \begin{equation}\label{eq:pred_r}
\hat{r}_{ai}=\mathbf{v}^T_i\mathbf{u}_a,
\end{equation}

\noindent where $\mathbf{u}_a$ is the embedding of $a$, which is the a-th column of  the user embedding matrix {\small$\mathbf{U}$}. Similarly, $\mathbf{v}_i$ represents item $i$'s embedding in the $i$-th column of item embedding matrix {\small$\mathbf{V}$}.

SVD++ is an enhanced version of the latent factor based models that leveraged the rated history items of each user for better user embedding modeling~\cite{KDD2008factorization}. In SVD++, each user's embedding is composed of a free embedding as classical latent factor based models, as well as an auxiliary embedding that is summarized from her rated items. Therefore, the predicted preference is modeled as:

\vspace{-0.2cm}
 \begin{equation}\label{eq:pred_svd++}
\hat{r}_{ai} = \mathbf{v}^T_i(\mathbf{u}_a+ \frac{1}{|R_a|}\sum_{j\in R_a}\mathbf{y}_j)
\end{equation}
\vspace{-0.2cm}

\noindent where {\small$R_a=[j: r_{aj}=1]$} is the itemset that $a$ shows implicit feedback, and $\mathbf{y}_j$ is an implicit factor vector.

As sometimes users and items are associated with attributes, the feature enriched embedding models give the predicted preference $r_{ai}$ of user $a$ to item $i$ is:

\vspace{-0.2cm}
 \begin{equation}\label{eq:pred_fm}
\hat{r}_{ai} =\mathbf{w}^T \mathbf{[\mathbf{x}_a, \mathbf{y}_i]}+\mathbf{v}^T_i\mathbf{u}_a,
\end{equation}

\noindent where the first term captures  the bias terms with the feature engineering, and the second term models the second-order interaction between users and items. 
Different embedding based models vary in the embedding matrix formulation and the optimization function. E.g.,  Bayesian Personalized Ranking~(BPR) is one of the most successful pair-wise based optimization function for implicit feedback~\cite{UAI2009bpr}. In BPR, it assumes that the embedding matrices {\small$\mathbf{U}$} and {\small$\mathbf{V}$} are free embeddings that follow a Gaussian prior, which is equivalent to adding a L2-norm regularization in the optimization function as:

\begin{small}
\vspace{-0.2cm}
\begin{equation}\label{eq:loss_BPR}
\min\limits_{[\mathbf{W},\mathbf{U},\mathbf{V}]} \mathcal{L}=\sum_{a=1}^M\sum\limits_{(i,j)\in D_a }\sigma(\hat{r}_{ai}-\hat{r}_{aj}) +\lambda(||\mathbf{U}||^2_F+||\mathbf{V}||_F)^2
\end{equation}
\vspace{-0.2cm}
\end{small}

\noindent where $\sigma(x)=\frac{1}{1+exp(-x)}$ is a logistic function that transforms the input into range $(0,1)$. {\small$D_a=\{(i,j)|i\in R_a\!\wedge\!j\in V-R_a\}$} is the training data for $a$ with {\small$R_a$} the itemset that $a$ positively shows feedback, and $j\in V-R_a$ denotes  the items that $a$ does not show feedback in the training data.

\textbf{Social Recommendation Models}
Social influence occurs when a persons's emotions, opinions or behaviors are affected by others~\cite{si_wiki}.
In a social platform, social scientists have converged that social influence is a natural process for users to disseminate their preferences to the followers in the social network, such that the action~(preference, interest) of a user changes with the influence from his/her trusted users~\cite{PNAS2012social,ibarra1993power,kempe2003maximizing,PNAS2014experimental}. Therefore, as the social diffusion process continues, the \emph{social correlation} phenomenon exits, with each user's  preference and behavior are similar to her social connections~\cite{PNAS2012social,KDD2008influence}.

The social influence and social correlation among users' interests are the foundation for building social recommender systems~\cite{WSDM2011recommender,SIGIR2018attentive,guo2015trustsvd}. Due to the superiority of embedding based models for recommendation, most social recommendation models are also built on these embedding models. These social embedding models could be summarized into the following two categories: the social regularization based approaches~\cite{Recsys2010matrix,WSDM2011recommender,TKDE2014scalable} and the user behavior enhancement based approaches~\cite{guo2015trustsvd,TKDE2016novel}. Specifically, the social regularization based approaches assumed that connected users would  show similar embeddings under the social influence diffusion. As such, besides the classical collaborative filtering based loss function~(e.g, Eq.\eqref{eq:loss_BPR}), an additional social regularization term is incorporated in the overall optimization function as:

\begin{small}
\vspace{-0.2cm}
\begin{equation}
\sum_{i=1}^M\sum_{j=1}^M s_{ij}||\mathbf{u}_i-\mathbf{u}_j||_F^2=\mathbf{U}(\mathbf{D}-\mathbf{S})\mathbf{U}^T,
\end{equation}
\vspace{-0.2cm}
\end{small}

\noindent where {\small $\mathbf{D}$} is a diagonal matrix with  $d_{aa}=\sum_{b=1}^M s_{bb}$.

Instead of the social regularization term, some researchers argued that the social network provides valuable information to enhance each user's behavior to alleviate the data sparsity issue. TrustSVD is such a model that shows state-of-the-art performance~\cite{guo2015trustsvd,TKDE2016novel}.
As researchers have well recognized that each user shows similar preferences as their social connections,  the implicit feedbacks of a user's social neighbors' on items could be regarded as the auxiliary feedback of this user, with the modeling process as:

\vspace{-0.2cm}
\begin{equation}\label{eq:trustsvd}
\hat{r}_{ai}=\mathbf{v}^T_i (\mathbf{u}_a+\sum_{b\in S_a}\frac{\mathbf{u}_b}{|S_a|})
\end{equation}
\vspace{-0.2cm}

\noindent where $\mathbf{u}_b$ denotes the latent embedding of user $b$, who is trusted by $a$. As such, $a$'s latent embedding is enhanced by considering the influence of her trusted users' latent embeddings in the social network.

Despite the performance improvement of incorporating the social network for social recommendation, we notice that nearly all of the current social recommendation models leveraged the observed social connections~(each user's social neighbors) for recommendation with a static process at once.
However, the social influence is not a static but a recursive process, with each user is influenced by the social connections as time goes on. At each time, users need to balance their previous preferences with the influences from social neighbors to form their updated latent interests. Then, as the current user interest evolves, the influences from social neighbors changes. The process is recursively diffused in the social network. Therefore, current solutions neglected the iterative social diffusion process for social recommendation. What's worse, when the user features are available, these social recommendation models need to be redesigned to leverage the feature data for better correlation modeling between users~\cite{TKDE2014scalable}.

\begin{small}
\begin{figure*} [htb]
	\begin{center}
		\includegraphics[width=170mm]{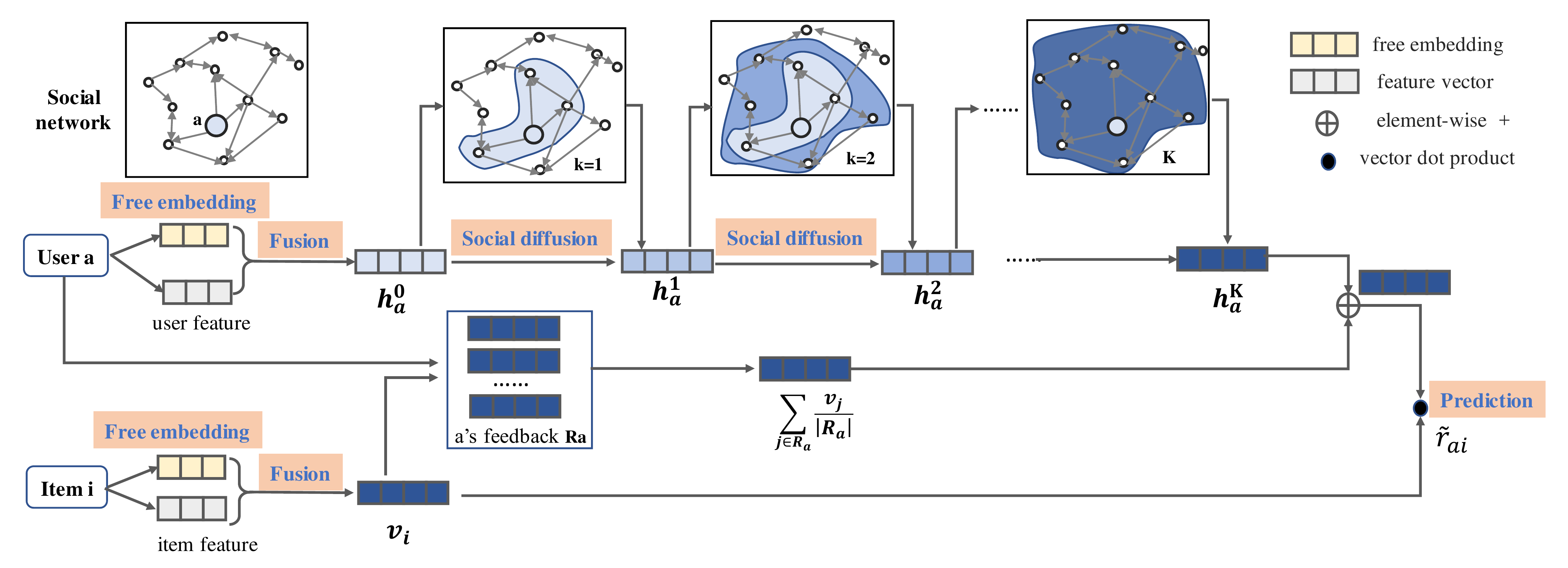}
	\end{center}
    \vspace{-0.5cm}
	\caption{The overall architecture of our proposed model. The four parts of DiffNet are shown with orange background.}\label{fig:framework}
   \vspace{-0.3cm}
\end{figure*}
\end{small}

\section{The Proposed Model }
In this part, we build a DiffNet model that stimulates the influence diffusion for social recommendation. We start with the overall architecture of DiffNet, followed by the model learning process. Finally, we give a detailed discussion of the proposed model.

\subsection{Model Architecture}
We show the overall neural architecture of DiffNet in Fig~\ref{fig:framework}. By taking an user-item pair $<a,i>$ as input, it outputs  the probability  $\hat{r}_{ai}$ that $u$ would like item $i$.  The overall neural architecture of DiffNet  contains four main parts: the embedding layer, the fusion layer, the layer-wise influence diffusion layers, and the prediction layer.  Specifically, by taking related inputs, the embedding layer outputs free embeddings  of users and items. For each user~(item), the fusion layer generates a hybrid user~(item) embedding by fusing both a user's~( an item's) free embedding  and the associated features. The fused user embedding is then sent to the influence diffusion layers. The influence diffusion layers are built with a layer-wise structure to model the recursive social diffusion process in the social network, which is the key idea of the DiffNet. After the influence diffusion process reaches stable, the output layer generates the final predicted preference of a user-item pair. We detail each part as follows:

%
%
%
%

\textbf{Embedding Layer.} Similar as many embedding based recommendation models~\cite{UAI2009bpr,rendle2012factorization,WWW2017neural}, let {\small$\mathbf{P}\in\mathbb{R}^{D\times M}$} and {\small$\mathbf{Q}\in\mathbb{R}^{D\times N}$} represent the free embeddings of users and items. These free embeddings capture the collaborative latent representations of users and items.  Given the one hot representations of user $a$ and item $i$, the embedding layer performs an index operation and outputs the free user latent vector $\mathbf{p}_a$ and free item latent vector $\mathbf{q}_i$ from user free embedding matrix {\small$\mathbf{P}$} and item free embedding matrix {\small$\mathbf{Q}$}.

\textbf{Fusion Layer.} For each user $a$, the fusion layer takes $\mathbf{p}_a$ and her associated feature vector $\mathbf{x}_a$ as input, and outputs a user fusion embedding $\mathbf{h}^0_a$ that captures the user's initial interests from different kinds of input data. We model the fusion layer as a one-layer fully connected neural network as:


\begin{small}
\begin{equation}\label{eq:fusion_user}
\mathbf{h}^0_a = g(\mathbf{W}^0\times[\mathbf{x}_a,\mathbf{p}_a]),
\end{equation}
\end{small}

\noindent where $\mathbf{W}^0$  is a transformation matrix, and $g(x)$ is a non-linear function.  Without confusion, we omit the bias term in a fully-connected neural network for notational convenience. This fusion layer could generalize many typical fusion operations, such as the concatenation  operation as $\mathbf{h}^0_a=[\mathbf{x}_a, \mathbf{p}_a]$  by setting {\small$\mathbf{W}^0$} as an identity matrix.

Similarly, for each item $i$, the fusion layer models the  item embedding  $\mathbf{v}_i$ as a non-linear transformation between its free latent vector $\mathbf{q}_i$ and its feature vector $\mathbf{y}_i$ as:

\vspace{-0.2cm}
\begin{equation} \label{eq:item_embed}
\mathbf{v}_i = \sigma(\mathbf{F}\times[\mathbf{q}_i, \mathbf{y}_i]).
\end{equation}
\vspace{-0.2cm}


\textbf{Influence Diffusion Layers.} By feeding  the output of each user $a$'s fusion embedding $\mathbf{h}^0_a$ from the fusion layer into the influence diffusion part, the influence diffusion layers model the dynamics of users' latent preference diffusion in the social network $\mathbf{S}$.   As information diffuses in the social network from time to time, the influence diffusion part is analogously built with a multi-layer structure. Each layer $k$ takes the users' embeddings from the previous layers as input, and output users' updated embeddings after the current social diffusion process finishes. Then, the updated user embeddings are sent to the $k+1$-th layer for the next diffusion process.

For each user $a$,  let $\mathbf{h}^{k}_a$ denote her latent embedding in the $k$-th layer of the influence diffusion part. By feeding the output of the $k$-th layer into the $k+1$-th layer, the  influence diffusion operation at the $k+1$-th social diffusion layer updates each user $a$'s latent embedding into $\mathbf{h}_a^{k+1}$. Specifically, the updated embedding $\mathbf{h}_a^{k+1}$  is composed of two steps: diffusion influence aggregation~(AGG) from $a$'s trusted users from the $k$-th layer, which transforms all the social trusted users' influences into a fixed length vector $\mathbf{h}^{k+1}_{Sa}$:

\begin{equation}\label{eq:pro_gcn1}
\mathbf{h}^{k+1}_{Sa}=Pool(\mathbf{h}^k_b| b\in S_a),
\end{equation}

\noindent where the $Pool$ function could be defined as an average pooling that performs a mean operation of
all the trusted users' latent embedding at the $k$-th layer. The $Pool$ can also be defined as a max operation that select the maximum element of all the trusted users' latent embedding at the $k$-th layer to form $\mathbf{h}^{k+1}_{Sa}$.

Then, $a$-th updated embedding $\mathbf{h}^{(k+1)}_a$ is a combination of her latent embedding $\mathbf{h}^k_a$ at the $k$-th layer and the influence diffusion embedding aggregation $\mathbf{h}^{k+1}_{Sa}$ from her trusted users. Since we do not know how each user balances these two parts, we use a non-linear neural network to model the combination as:

\begin{small}
\begin{equation}\label{eq:pro_gcn2}
\mathbf{h}^{k+1}_a=s^{(k+1)}(\mathbf{W}^k\times[\mathbf{h}^{k+1}_{S_a}, \mathbf{h}^k_a]),
\end{equation}
\end{small}

\noindent where $s^k(x)$ is non-linear transformation function.

With a predefined diffusion depth $K$, for each user $a$, the influence diffusion layer starts with the  layer-0 user embedding $\mathbf{h}^0_a$~(Eq.\eqref{eq:fusion_user}), i.e., the output of the fusion layer, and the layer-wise influence diffusion process then diffuses to layer 1, followed by layer 1 diffuses to layer 2. This influence diffusion step is repeated for $K$ steps to reach the diffusion depth $K$, where each user $a$'s
latent embedding at the $K$-th layer is $\mathbf{h}^K_a$.

Please note that, DiffNet only diffuses users' latent vectors in the influence diffusion part without any item vector diffusion modeling. This is quite reasonable as item latent embeddings are static and do not propagate in the social network.

\textbf{Prediction Layer.} Given each user $a$'s embedding  $\mathbf{h}^K_a$ at the $K$-th layer after the iterative diffusion process, each item $i$'s fusion vector  $\mathbf{v}_i$, we model the predicted preference of user $a$ to item $i$ as:

\begin{small}
\vspace{-0.2cm}
\begin{flalign}
\label{eq:pred_dip1}&\mathbf{u}_a =\mathbf{h}^K_a+\sum_{i\in R_a} \frac{\mathbf{v}_i}{|R_a|}, \\
\label{eq:pred_dip} &\hat{r}_{ai}=\mathbf{v}^T_i\mathbf{u}_a,
\end{flalign}
\vspace{-0.4cm}
\end{small}

\noindent where $R_a$ is the itemset that $a$ likes. In this equation, each user's final latent representation $\mathbf{u}_a$ is composed of two parts: the embeddings from the output of the social diffusion layers as $\mathbf{h}^K_a$, and the preferences from her historical behaviors as: $\sum_{i\in R_a} \frac{\mathbf{v}_i}{|R_a|}$. Specifically, the first term captures the user's interests from the recursive social diffusion process in the social network structure.  The second term resembles the SVD++ model that leveraged the historical feedbacks of the user to alleviate the data sparsity of classical CF models~\cite{KDD2008factorization}, which has shown better performance over the classical latent factor based models.
Thus, the final user embedding part is more representative with the recursive social diffusion modeling and the historical feedbacks of the user. After that, the final predicted rating is still measured by the inner produce between the corresponding user final latent vector and item latent vector.

\subsection{Model Training}
As we focus on implicit feedbacks of users, similar to the widely used ranking based loss function in BPR~\cite{UAI2009bpr}, we also design a pair-wise ranking  based loss function for optimization:

\begin{small}
	\begin{equation}\label{eq:loss_r}
	\min\limits_{\Theta} \mathcal{L}(\mathbf{R},\mathbf{\hat{R}})=\sum_{a=1}^M\sum\limits_{(i,j)\in D_a } \sigma(\hat{r}_{ai}-\hat{r}_{aj}) +\lambda||\Theta_1||^2
	\end{equation}
\end{small}

\noindent where $\sigma(x)$ is a sigmoid function. {\small $\Theta\!=\![\Theta_1,\Theta_2]$}, with {\small $\Theta_1\!=\![\mathbf{P},\mathbf{Q}]$}, and
{\small $\Theta_2\!=\![\mathbf{F}, {[\mathbf{W}^k]}_{k=0}^{K-1}]$}.  $\lambda$ is a regularization parameter that controls the complexity of user and item free embedding matrices. {\small$D_a=\{(i,j)|i\in R_a\!\wedge\!j\in V-R_a\}$} denotes the pairwise training data for $a$ with {\small$R_a$} represents the itemset that $a$ positively shows feedback.

All the parameters in the above loss function are differentiable. In practice, we implement the proposed model with TensorFlow\footnote{https://www.tensorflow.org} to train model parameters with mini-batch Adam. We split the mini-batch according to the userset, i.e., each user's training records are ensured in the same mini-batch. This mini-batch splitting procedure avoids the repeated computation of each user $a$'s latent embedding $\mathbf{h}^K_a$ in the iterative influence diffusion layers. 

As we could only observe positive feedbacks of users with huge missing unobserved values, similar as many implicit feedback works, for each positive feedback, we randomly sample 10 missing unobserved feedbacks as pseudo negative feedbacks at each iteration in the training process~\cite{leAAAI2015}.  As each iteration the pseudo negative samples change, each missing value gives very weak negative signal.

\subsection{Discussion}

\subsubsection{Complexity}

\textbf{Space complexity.}
As shown in Eq.\eqref{eq:loss_r}, the model parameters are composed of two parts: the user and item free embeddings {\small$\Theta_1\!=\![\mathbf{P},\mathbf{Q}]$}, and the parameter set {\small $\Theta_2\!=\![\mathbf{F}, {[\mathbf{W}^k]}_{k=0}^{K-1}]$}. Since most embedding based models~(e.g., BPR~\cite{UAI2009bpr}, FM~\cite{rendle2012factorization}) need to store the embeddings of each user and each item, the space complexity of $\Theta_1$ is the same as classical embedding based models and grows linearly with users and items. For parameters in {\small$\Theta_2$}, they are shared among all users and items,  with the dimension of each parameter is far less than the number of users and items. This additional storage cost is a small constant that could be neglected. Therefore, the space complexity of DiffNet is the same as classical embedding models.

\textbf{Time complexity.} Since our proposed loss function resembles the BPR model with the pair-wise loss function that is designed for implicit feedback, we compare the time complexity of DiffNet with BPR. The main additional time cost lies in the layer-wise influence diffusion process. The dynamic diffusion process costs {\small$O(MKL)$}, where $M$ is the number of users, and  $K$ denotes the diffusion depth and $L$ denotes the average social neighbors of each user. Similarly, the additional time complexity of updating parameters is {\small$O(MKL)$}. Therefore, the additional time complexity is {\small$O(MKL)$}. In fact, as shown in the empirical findings as well as our experimental results, DiffNet reaches the best performance when {\small$K=2$}. Also, the average social neighbors per user are limited with {\small $L\ll M$}. Therefore, the additional time complexity is acceptable and the proposed DiffNet could be applied to real-world social recommender systems.

\subsubsection{Model Generalization.} The proposed DiffNet model is designed under the problem setting with the input of  user feature matrix {\small$\mathbf{X}$},  item feature matrix {\small$\mathbf{Y}$}, and the social network {\small$\mathbf{S}$}. Specifically, the fusion layer takes users'~(items') feature matrix   for user~(item) representation learning. The layer-wise diffusion layer utilized the social network structure {\small$\mathbf{S}$} to model how users' latent preferences are dynamically influenced from the recursive social diffusion process. Next, we would show that our proposed model is generally applicable when different kinds of data input are not available.


When the user~(item) features are not available, the fusion layer disappears. In other words, as shown in Eq.\eqref{eq:item_embed}, each item's latent embedding $\mathbf{v}_i$ degenerates to $\mathbf{q}_i$. Similarly, each user's initial layer-0 latent embedding  $\mathbf{h}^0\!=\!\mathbf{p}_a$ ~(Eq.\eqref{eq:fusion_user}).  Similarly, when either the user attributes or the item attributes do not exist, the corresponding fusion layer of user or item degenerates.  

The key idea of our proposed model is the carefully designed social diffusion layers with the input social network {\small$\mathbf{S}$}. When the recommender system does not contain any social network information, the social diffusion layers disappear with $\mathbf{h}^K_a=\mathbf{h}^0_a$. Under this circumstances, as shown in Eq.~\eqref{eq:pred_dip} our proposed model degenerates to an enhanced SVD++ model~\cite{KDD2008factorization} for recommendation, with the user and item latent embeddings contain the fused free embeddings and the associated user and item features.

\subsubsection{Comparisons to Graph Convolutional based Models}

In our proposed DiffNet, the designed layer-wise diffusion part~(Eq.\eqref{eq:pro_gcn1} and Eq.\eqref{eq:pro_gcn2}) presents similar idea as the  Graph Convolutional Networks~(GCN), which are state-of-the-art representation learning techniques of graphs ~\cite{IDEB2017representation,ICLR2017semi,ICLR2017graph}.
GCNs generate node embeddings in a recursive message passing or information diffusion manner of a graph, where the representation vector of a node is computed recursively from aggregation features in neighbor nodes. GCNs has shown theoretical elegance as simplified version of spectral based graph models~\cite{ICLR2017semi}. Besides, GCNs are time efficient and achieve better performance in many graph-based tasks.

Due to the success of GCNs, several models have attempted to transfer the idea of GCNs for the recommendation tasks. By transferring these models to the recommendation scenario, the main components are how to construct a graph and further exploit the uniqueness of recommendation properties. Among them, the most closely related works are GC-MC~\cite{ICLR2017graph} and PinSage~\cite{ICLR2017graph}.

\textbf{GC-MC:} It is one of the first few attempts that directly applied the graph convolutions for recommendation. GC-MC defines a user-item bipartite graph from user-item interaction behavior~\cite{ICLR2017graph}. Then, each user embedding is convolved as the aggregation of the embeddings of her rated items. Similarly, each item embedding is convolved as the  aggregation of the embeddings of the rated users' embeddings. However, the graph convolution is only operated with one layer of the observed links between users and items, neglecting the layer-wise diffusion structure of the graph.

\textbf{PinSage:} It is designed for similar item recommendation from a large recommender  system.
By constructing an item-item correlation graph from users' behaviors, a data-efficient GCN algorithm PinSage is developed~\cite{KDD2018graph}. PinSage could incorporate both the item correlation graph as well as node features to generate item embeddings. The main contribution lies in how to design efficient sampling techniques to speed up the training process. Instead of message passing on item-item graph, our work performs the recursive information diffusion of the social network, which is more realistic to reflect how users are dynamically influenced by the social influence diffusion. Applying GCNs for social recommendation is quite natural and to the best of our knowledge, has not been studied before.


\section{Experiments}
In this section, we conduct experiments to evaluate the performance of DiffNet on two datasets. Specifically, we aim to answer the following two research questions: First, does DiffNet outperforms the state-of-the-art baselines for the social recommendation task? Second, what is the performance of these models under different data sparsity?  Third, the effectiveness of each part in DiffNet, e.g., diffusion modeling, attributes modeling, and so on.

%



\begin{small}
\begin{table}
	\centering
	\setlength{\belowcaptionskip}{5pt}
    \vspace{-0.5cm}
	\caption{The statistics of the two datasets.}\label{tab:stat}
\vspace{-0.3cm}
	\begin{tabular}{c|c|c}
		\hline
		Dataset&\textit{Yelp}&\textit{Flickr}\\
		\hline
		Users&17237&8358\\
		Items&38342&82120\\
		\hline
		Total Links&143765&187273\\
        Ratings &204448 & 314809 \\
		\hline
		Link Density&0.048\%&0.268\%\\
		Rating Density&0.031\%&0.046\%\\
		\hline
	\end{tabular}
\end{table}
\vspace{-0.3cm}
\end{small}

\begin{table*}
	\begin{small}
		\centering
        \vspace{-0.3cm}
		\caption{HR@10 and NDCG@10 comparisons for different dimension size $D$.}\label{tab:hr_ndcg_d}
        \vspace{-0.3cm}
		\begin{tabular}{|c|c|c|c|c|c|c|c|c|c|c|c|c|}
			\hline
			\multirow{3}*{Models} &\multicolumn{6}{|c|}{\textit{Yelp}}&\multicolumn{6}{|c|}{\textit{Flickr}} \\
			\cline{2-13}
			&\multicolumn{3}{|c|}{HR}&\multicolumn{3}{|c|}{NDCG}&\multicolumn{3}{|c|}{HR}&\multicolumn{3}{|c|}{NDCG}\\
			\cline{2-13}
			&$D$=16&$D$=32&$D$=64&$D$=16&$D$=32&$D$=64&$D$=16&$D$=32&$D$=64&$D$=16&$D$=32&$D$=64\\
			\hline
			\hline
			BPR&0.2443&0.2632&0.2617&0.1471&0.1575&0.155&0.0851&0.0832&0.0791&0.0679&0.0661&0.0625\\
			\hline
            SVD++&0.2581&0.2727&0.2831&0.1545&0.1632&0.1711&0.0821&0.0934&0.1054&0.0694&0.0722&0.0825\\
            \hline
			FM&0.2768&0.2835&0.2825&0.1698&0.1720&0.1717&0.1115&0.1212&0.1233&0.0872&0.0968&0.0954\\
			\hline
			TrustSVD&0.2853&0.2880&0.2915&0.1704&0.1723&0.1738&0.1372&0.1367&0.1427&0.1062&0.1047&0.1085\\
			\hline
			ContextMF&0.2985&0.3011&0.3043&0.1758&0.1808&0.1818&0.1405&0.1382&0.1433&0.1085&0.1079&0.1102\\
			\hline
            GC-MC &0.2876&0.2902&0.2937&0.1657&0.1686&0.174&0.1123&0.1155&0.1182&0.0883&0.9450&0.0956\\
            \hline
			PinSage&0.2952&0.2958&0.3065&0.1758&0.1779&0.1868&0.1209&0.1227&0.1242&0.0952&0.0978&0.0991\\
			\hline
			DiffNet&\textbf{0.3366}&\textbf{0.3437}&\textbf{0.3477}&\textbf{0.2052}&\textbf{0.2095}&\textbf{0.2121}&\textbf{0.1575}&\textbf{0.1621}&\textbf{0.1641}&\textbf{0.1210}&\textbf{0.1231}&\textbf{0.1273}\\
			\hline
		\end{tabular}
	\end{small}
\vspace{-0.3cm}
\end{table*}

\begin{table*}
	\begin{small}
		\centering
		\caption{HR@N and NDCG@N comparisons for different top-N values.}\label{tab:hr_ndcg_topk}
        \vspace{-0.3cm}
		\begin{tabular}{|c|c|c|c|c|c|c|c|c|c|c|c|c|}
			\hline
			\multirow{3}*{Models} &\multicolumn{6}{|c|}{\textit{Yelp}}&\multicolumn{6}{|c|}{\textit{Flickr}} \\
			\cline{2-13}
			&\multicolumn{3}{|c|}{HR}&\multicolumn{3}{|c|}{NDCG}&\multicolumn{3}{|c|}{HR}&\multicolumn{3}{|c|}{NDCG}\\
			\cline{2-13}
			&N=5&N=10&N=15&N=5&N=10&N=15&N=5&N=10&N=15&N=5&N=10&N=15\\
			\hline
			\hline
			BPR&0.1713&0.2632&0.3289&0.1243&0.1575&0.1773&0.0657&0.0851&0.1041&0.0607&0.0679&0.0737\\
			\hline
            SVD++&0.1868&0.2831&0.3492&0.1389&0.1711&0.1924&0.0827&0.1054&0.1257&0.0753&0.0825&0.0895\\
			\hline
			FM&0.1881&0.2835&0.3463&0.1359&0.1720&0.1895&0.0918&0.1233&0.1458&0.0845&0.0968&0.1046\\
			\hline
			TrustSVD&0.1906&0.2915&0.3693&0.1385&0.1738&0.1983&0.1072&0.1427&0.1741&0.0970&0.1085&0.1200\\
			\hline
			ContextMF&0.2045&0.3043&0.3832&0.1484&0.1818&0.2081&0.1095&0.1433&0.1768&0.0920&0.1102&0.1131\\
			\hline
            GC-MC&0.1932&0.2937&0.3652&0.1420&0.1740&0.1922&0.0897&0.1182&0.1392&0.0795&0.0956&0.1002\\
			\hline
			PinSage&0.2099&0.3065&0.3873&0.1536&0.1868&0.2130&0.0925&0.1242&0.1489&0.0842&0.0991&0.1036\\
			\hline
			DiffNet&\textbf{0.2276}&\textbf{0.3477}&\textbf{0.4232}&\textbf{0.1679}&\textbf{0.2121}&\textbf{0.2331}&\textbf{0.1210}&\textbf{0.1641}&\textbf{0.1952}&\textbf{0.1142}&\textbf{0.1273}&\textbf{0.1384} \\
			\hline
		\end{tabular}
	\end{small}
\vspace{-0.3cm}
\end{table*}

\subsection{Experimental Settings}

\textbf{Datasets.}
\textit{Yelp} is an online location-based social network. Users make friends with others and express their experience through the form of reviews and ratings. As each user give ratings in the range $[0,5]$, similar to many works,  we transform the ratings that are larger than 3 as the liked items by this user. As the rich reviews are associated with users and items, we use the
popular gensim tool\footnote{https://radimrehurek.com/gensim/} to learn the embedding of each word with Word2vec model~\cite{mikolov2013distributed}. Then, we get the feature vector of each user~(item) by averaging all the learned word vectors of the user(item).

\textit{Flickr} is a who-trust-whom online image based social sharing platform. Users follow other users and share their preferences to images to their social followers. Users express their preferences through the upvote behavior. For research purpose, we crawl a large dataset from this platform. Given each image, we have a ground truth classification of this image on the dataset. We send images to a VGG16 convolutional neural network and treat the 4096 dimensional representation in the last connected layer in VGG16 as the feature representation of the image~\cite{simonyan2014very}. For each user, her feature representation is the average of the image feature representations she liked in the training data.

In the data preprocessing step, for both datasets, we filtered out users that have less than 2 rating records and 2 social links, and removed the items which have been rated less than 2 times. We randomly select 10\% of the data for the test. In the remaining 90\% data, to tune the parameters, we select 10\% from the training data as the validation set. The detailed statistics of the data after preprocessing is shown in Table~\ref{tab:stat}.

\textbf{Baselines and Evaluation Metrics.}
We compare DiffNet with various state-of-the-art baselines, including the classical pair-wise based recommendation model \emph{BPR}~\cite{UAI2009bpr}, feature enhanced latent factor model \emph{FM}~\cite{ICDM2010factorization}, a state of the art social recommendation model \emph{TrustSVD}~\cite{guo2015trustsvd}, a context-aware social recommendation model \emph{ContextMF} that utilized the same input as our proposed model for recommendation~\cite{TKDE2014scalable}. Besides, we also compare our proposed model with two graph convolutional based recommendation models: \emph{GC-MC}~\cite{ICLR2017graph} and \emph{PinSage}~\cite{KDD2018graph}. As the original PinSage focuses on generating high-quality embeddings of items, we generalize this model by constructing a user-item bipartite for recommendation~\cite{KDD2018graph}. Both of these two convolutional recommender models utilized the user-item bipartite and the associated features of users and items for recommendation.

As we focus on recommending top-N items for each user, we use two widely adopted ranking based metrics: Hit Ratio~(HR) and Normalized Discounted Cumulative Gain(NDCG)~\cite{SIGIR2018attentive}. Specifically, HR measures the number of items that the user likes in the test data that has been successfully predicted in the top-N ranking list. And NDCG considers the hit positions of the items and gives a higher score if the hit items in the top positions.
For both metrics, the larger the values, the better the performance.
Since there are too many unrated items, in order to reduce the computational cost, for each user, we randomly sample 1000 unrated items at each time and combine them with the positive items the user likes in the ranking process. We repeat this procedure 10 times and report the average ranking results.


\textbf{Parameter Setting.} For all the models that are based on the latent factor models, we initialize the latent vectors with small random values. In the model learning process, we use Adam as the optimizing method for all models that relied on the gradient descent based methods with an initial learning rate of 0.001. And the batch size is set as 512. In our proposed DiffNet model, we try the regularization parameter $\lambda$ in the range $[0.0001,0.001,0.01,0.1]$, and find $\lambda=0.001$ reaches the best performance. For the aggregation function in Eq.\eqref{eq:pro_gcn1}, we have tried the max pooling and average pooling. We find the average pooling usually shows better performance. Hence, we set the average pooling as the aggregation function. Similar to many GCN models~\cite{KDD2018graph,ICLR2017semi}, we set the depth parameter $K=2$. With the user and item free embedding size $D$, in the fusion layer and the following influence diffusion layers, each layer's output is also set as $D$ dimension. For the non-linear function $g(x)$ in the fusion layer, we use a sigmoid function that transforms each value into range $(0,1)$. And we set the non linear functions  of $[s^k(x)]_{k=0}^{K-1}$ with the ReLU function to avoid the vanishing gradient problem. After the training of each layer, we use batch normalization to avoid the internal covariate shift problem~\cite{ioffe2015batch}.  There are several parameters in the baselines, we tune all these parameters to ensure the best performance of the baselines for fair comparison. Please note that as generating user and item features are not the focus of our paper, we use the feature construction techniques as mentioned above. However, our proposed model can be seamlessly incorporated with more advanced feature engineering techniques.

\subsection{Overall Comparison}
In this section, we compare the overall performance of all models on two datasets. Specifically, Table \ref{tab:hr_ndcg_d} shows the HR@10 and NDCG@10 results for both datasets with varying latent dimension size $D$.
Among all the baselines, BPR only considered the user-item rating information for recommendation, FM and TrustSVD improve over BPR by leveraging the node features and social network information. PinSage takes the same kind of input as FM and shows better performance than FM, showing the effectiveness of modeling the information passing of a graph. ContextMF is the baseline that uses the user and item features, as well as the social network structure. It performs better than most baselines. Our proposed model consistently outperforms ContextMF, showing the effectiveness of modeling the recursive social diffusion process in the social recommendation process.

When comparing the results of the two datasets, we observe that leveraging the social network  structure and the social diffusion  process contributes more on \textit{Flickr} compared to \textit{Yelp}. On both datasets, PinSAGE is the best baseline that leverages the node features without the social network information. E.g., DiffNet improves over PinSAGE about 13\% on \textit{Yelp}, and nearly 30\% on \textit{Flickr}. We guess a possible reason is that, the \textit{Flickr} is a social based image sharing platform with a stronger social influence diffusion effect.  In contrast, \textit{Yelp} is a location based social network, and users' food and shopping preferences are not easily influenced in the social platform. Last but not least,  we find the performance of all models does not increase as the latent dimension size $D$ increases from 16 to 64. Some models reach the best performance when $D=32$~(e.g., BPR) while other models reach the best performance when $D=64$~(e.g., DiffNet). In the following experiment, we set the proper $D$ for each model with the best performance in order to ensure fairness.

Table \ref{tab:hr_ndcg_topk} shows the HR@N and NDCG@N on both datasets with varying top-N recommendation size N. From the results, we also find similar observations as Table \ref{tab:hr_ndcg_d}, with our proposed model DiffNet always shows the best performance. Based on the overall experiment results, we could empirically conclude that our proposed DiffNet model outperforms all the baselines under different ranking metrics and different parameters.

\begin{figure*} [htb]
\vspace{-0.5cm}
  \begin{center}
  \vspace{-0.2cm}
      \subfigure[\vspace{-1.5cm}Yelp]{\includegraphics[width=80mm]{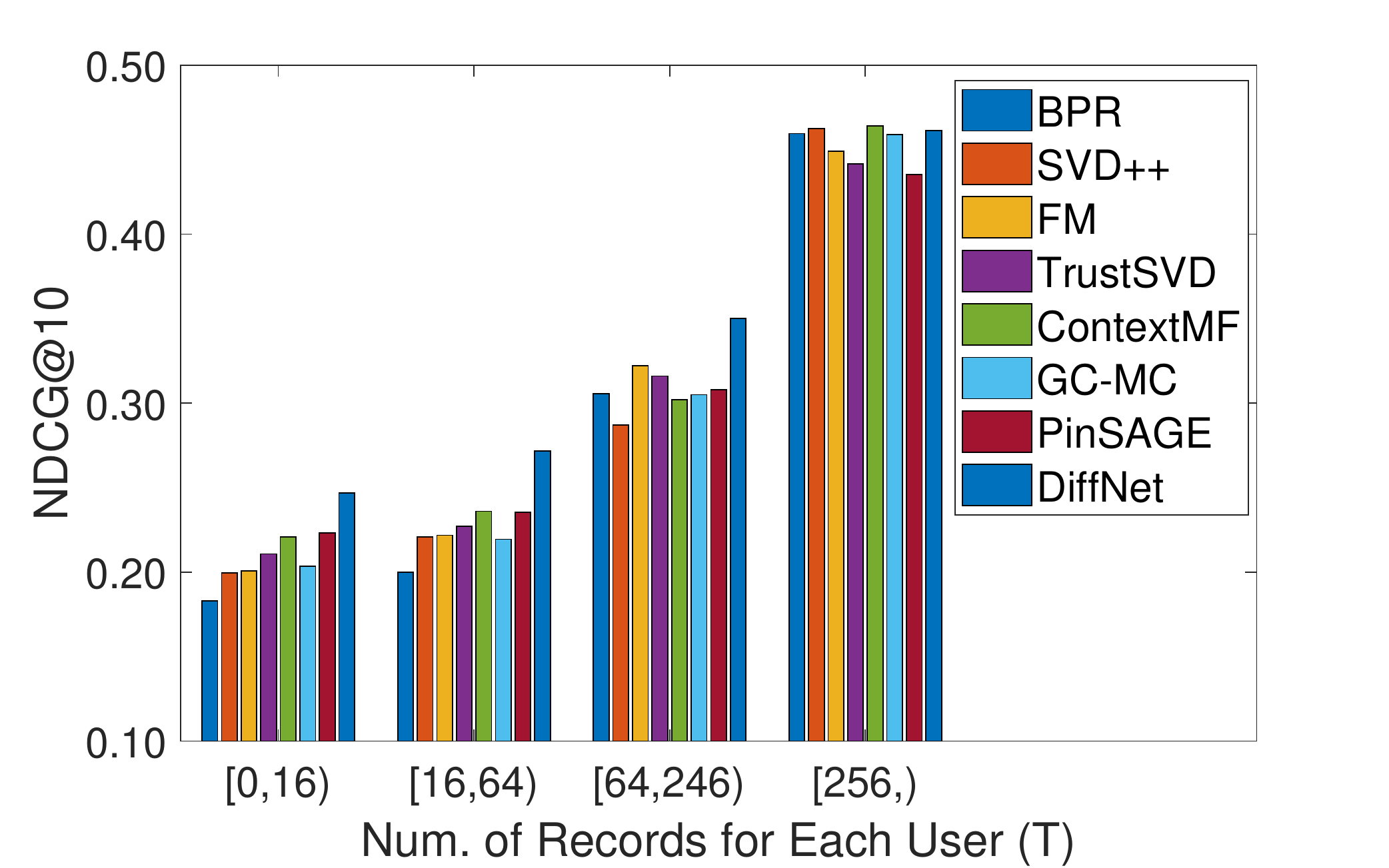} \label{fig:bin_user}}
      \subfigure[\vspace{-1.5cm}Flickr]{\includegraphics[width=80mm]{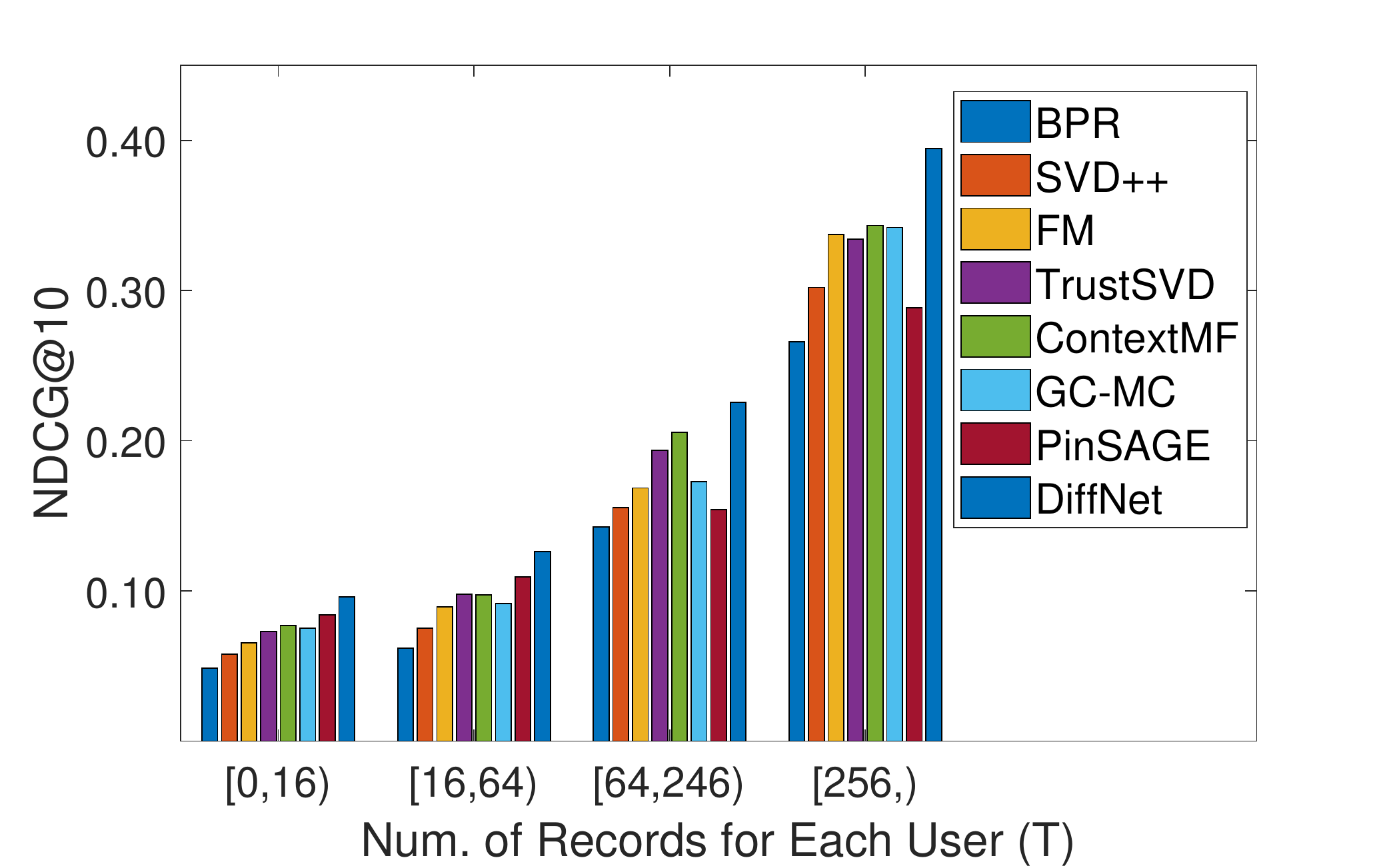}  \label{fig:bin_flickr}}
  \end{center}\vspace{-0.5cm}
  \caption{\small{Performance under different sparsity~(Better viewed in color.)}} \label{fig:bin_results}
  \vspace{-0.2cm}
\end{figure*}

\vspace{-0.2cm}
\subsection{Performance under Different Data Sparsity}
The data sparsity issue is a main challenge for most CF based recommender systems. 
In this part, we would like to show the performance of various models under different sparsity.

Specifically, we bin users into different groups based on the number of observed feedbacks in the training data. Then we show the performance of each group with different models on NDCG@10 in Fig~\ref{fig:bin_results}.  In this figure, the horizontal axis shows the user group information. E.g., $[16, 64)$ means for each user in this group, the training records satisfy $16\leq|R_a|<64$. As can be observed from this figure, for both datasets, with the increase of the user rating records, the performance increases quickly for all models. When the rating records of each user is less than 16, the BPR baseline could not work well as this model only exploited the very sparse user-item interaction behavior for recommendation. Under this situation, all improvement is significant by leveraging various kinds of side information. E.g., FM, SVD++, and ContextMF improves over BPR by 9.6\%, 15.2\% and 20.7\% on Yelp, and 34.9\%, 50.3\%, 58.4\% on Flickr. The improvement of Flick is much more significant than that of Yelp, as Flickr has much more items compared to Yelp. By considering the iteratively social diffusion process in social recommendation, our proposed model improves BPR by 34.8\% and 97.1\% on Flick and Yelp, which far exceeds the remaining models. With the increase of user rating records, the performance improvements of all models over BPR decrease, but the overall trend is that all models have better performance than BPR. We also observe that when users have more than 256 records, some methods have similar results as BPR or even a little worse than BPR. We guess a possible reason is that BPR could well learn the user interests from enough interaction data. Withe the additional side information, some noises are introduced to decrease the performance.

\begin{table*}
	\centering
	\caption{HR@10 and NDCG@10 performance with different diffusion depth $K$.}\label{tab:submodels_k}
	\begin{tabular}{|c|c|c|c|c|c|c|c|c|}
		\hline
		\multirow{2}*{Diffusion Depth $K$} &\multicolumn{4}{|c|}{\textit{Yelp}}&\multicolumn{4}{|c|}{\textit{Flickr}}\\
		\cline{2-9}
		&HR&Improve.&NDCG&Improve.&HR&Improve.&NDCG&Improve.\\
		\hline

		K=2    &\textbf{0.3477}&-&\textbf{0.2121}&-&\textbf{0.1641}&-&\textbf{0.1273}&-\\
		\hline
		K=0    &0.3145 &-9.54\%& 0.2014& -5.09\%& 0.1439&-12.27\%& 0.1148&-10.0\%\\
		\hline
		K=1    &0.3390& -2.50\%& 0.2981& -1.93\%&  0.1592 & -2.96\%& 0.1257&-1.22\%\\
		\hline
	    K=3   & 0.3348&-3.72\%& 0.2005&-5.49\%& 0.1603& -2.34\%&0.1246&-2.22\%\\
		\hline

	\end{tabular}
\end{table*}

\begin{table*}
	\centering
	\caption{HR@10 and NDCG@10 of our simplified models on \textit{Yelp} and \textit{Flickr} with different fusion inputs. \textbf{X}=\textbf{Y}=0 denotes the user and item feature vector are not available. \textbf{P}=0~(\textbf{Q}=0) denotes we do not add the free base user~(item) latent embedding.}\label{tab:submodels_y}
	\begin{tabular}{|c|c|c|c|c|c|c|c|c|}
		\hline
		\multirow{2}*{Simplified models} &\multicolumn{4}{|c|}{\textit{Yelp}}&\multicolumn{4}{|c|}{\textit{Flickr}}\\
		\cline{2-9}
		&HR&Improve.&NDCG&Improve.&HR&Improve.&NDCG&Improve.\\
		\hline
		DiffNet &\textbf{0.3477}&-&\textbf{0.2121}&-&\textbf{0.1641}&-&\textbf{0.1273}&-\\
        \hline
		$\mathbf{X}=0$  &0.3403&-2.11\%&0.2072&-2.32\%&0.1582&-3.58\%&0.1232&-3.25\%\\
		\hline
		$\mathbf{Y}=0$  &0.3271&-5.92\%&0.1951&-8.06\%&0.1423&-13.26\%&0.1098&-13.73\%\\
		\hline
	    $\mathbf{X}=\mathbf{Y}=0$ &0.3196&-8.09\%&0.1912& -9.89\% & 0.1360&-17.08\%&0.1073& -15.69\%\\
		\hline
		$\mathbf{P}=0$ &0.2461& -29.22\% &0.1569&-26.05\%&0.1056&-35.66\%&0.0863&-32.17\%\\
        \hline
        $\mathbf{Q}=0$&0.1975 &-43.19\% &0.658&-69.00\%&0.0334&-78.78\%&0.022&-82.13\%\\
		\hline
	\end{tabular}
\end{table*}

\subsection{Detailed Model Analysis}

We would analyze the recursive social diffusion depth $K$, and the impact of the fusion layer that combines the collaborative free embedding and associated entity feature vector.

Table~\ref{tab:submodels_k} shows the results on DiffNet with different $K$ values. The column of ``Improve'' show the
performance changes compared to the best setting of DiffNet, i.e., $K=2$. When $K=0$, the layer-wise diffusion part disappears, and our proposed model degenerates to an enhanced SVD++ with entity feature modeling. As can be observed from this figure, as we leverage the layer wise diffusion process from $K=0$ to $K=1$, the performance increases quickly for for both datasets. For both datasets, the best performance reaches with two recursive diffusion depth, i.e., $K=2$. When $K$ continues to increase to 3, the performance drops for both datasets. We hypothesis that, considering the $K$-step recursive social diffusion process resembles the k-hop neighbors of each user. Since the social diffusion diminishes with time and the distance between each user and the $k$-hop neighbors, setting $K$ with 2 is enough for social recommendation. In fact, other related studies  have empirically find similar trends, with the best diffusion size is set as $K=2$ or $K=3$~\cite{KDD2018graph,ICLR2017semi}.

Table~\ref{tab:submodels_k} shows the performance on DiffNet with different fusion inputs.  As can be seen from this figure, the performance drops when the user and~(or) item features are not available. We also notice that it is very important to add the free  latent embeddings of users and items in the modeling process. As can be observed from this figure, the performance drops very quickly when either the user free latent embedding matrix {\small$\mathbf{P}$} or the item free embedding matrix {\small$\mathbf{Q}$} are not considered. E.g., the performance drops about 80\% on Flickr when the item free embedding is not considered. The reason is that, the item~(user) latent factors could not be captured by the item~(user) features. Therefore, learning the collaborative effect between users and items with the free embeddings is very important for the recommendation task.

\section{Related Work}


\textbf{Collaborative Filtering.} Given an user-item rating  matrix {\small$\mathbf{R}$}, CF usually projected both users and items in a same low latent space for comparison~\cite{koren2009matrix,NIPS2008probabilistic}. In reality, compared to the explicit ratings, it is more common for users implicitly express their feedbacks through action or inaction, such as click, add to cart or consumption ~\cite{hu2008collaborative,UAI2009bpr}. Bayesian Personalized Ranking~(BPR) is a state-of-the-art latent factor based technique for dealing with implicit feedback. Instead of directly predicting each user's point-wise explicit ratings, BPR modeled the pair-wise preferences with the assumption that users prefer the observed implicit feedbacks compared to the unobserved ones~\cite{UAI2009bpr}. Despite the relatively high performance, the data sparsity issue is a barrier to the performance of these collaborative filtering models. To tackle the data sparsity issue, many models have been proposed by extending these classical CF models. E.g., SVD++ is proposed to combine users' implicit feedbacks and explicit feedbacks for modeling users' latent interests~\cite{KDD2008factorization}.  Besides, as users and items are associated with rich attributes, Factorization Machine~(FM) is such a unified model that leverages the user and item attributes in latent factor based models~\cite{kdd2015collaborative}. Recently, some deep learning based models have been proposed to tackle the CF problem. E.g., instead of assuming user and item are interacted as shallow linear interaction function between user and item latent vectors, NeuMF is proposed to model the complex interactions between user and item embeddings~\cite{WWW2017neural}. As sometimes the user and item features are sparse, many deep learning based models have been proposed to tackle how to model these sparse features~\cite{guo2017deepfm,kdd2015collaborative}. In contrast to these works, we do not consider the scenario of sparse features and put emphasis on how to model the recursive social diffusion process for social recommendation.

\textbf{Social Recommendation}  With the prevalence of online social platforms, social recommendation has emerged as a promising direction that leverages the social network among users to enhance recommendation performance~\cite{WSDM2011recommender,guo2015trustsvd,TKDE2014scalable}. In fact, social scientists have long converged that as information diffuses in the social networks, users are influenced by their social connections with the social influence theory, leading to the phenomenon of similar preferences among social neighbors~\cite{KDD2008influence,ibarra1993power,Nature2012,KDD2018deepinf}. Social regularization has been empirically proven effective for social recommendation, with the assumption that similar users would share similar latent preferences under the popular latent factor based models~\cite{Recsys2010matrix,WSDM2011recommender}.
SBPR model is proposed into the pair-wise BPR model with the assumption that users tend to assign higher ratings to the items their friends prefer~\cite{CIKM2014leveraging}. By treating the social neighbors' preferences as the auxiliary implicit feedbacks of an active user, TrustSVD~\cite{guo2015trustsvd,TKDE2016novel} is proposed to incorporate the trust influence from social neighbors on top of SVD++~\cite{KDD2008factorization}.
As items are associated with attribute information~(e.g., item description, item visual information), ContextMF is proposed to combine social context and social network under a collective matrix factorization framework with carefully designed regularization terms~\cite{TKDE2014scalable}. Social recommendation has also been extended with social circles~\cite{tkde2014personalized} and the temporal context~\cite{SIGIR2018attentive}.
Recently, the problem of how bridge a few overlapping users in the two domains of the social network domain and information domain for better recommendation has also been considered~\cite{wang2017item}.

Instead of simply considering the local social neighbors of each user, our work differs from these works in explicitly modeling the recursive social diffusion process to better model each user's latent preference in the global social network.

\textbf{Graph Convolutional Networks and Applications.} Our proposed model with recursive social diffusion process borrows the recent advances of graph convolutional networks~(GCN)~\cite{ICLR2017semi,ICLR2017graph,IDEB2017representation}. GCNs have shown success to extend the  convolution operation from the regular Euclidean domains to non-Euclidean graph domains. Spectral graph convolutional neural network based approaches provide localized convolutions in the spectral domain~\cite{ICLR2014spectral,NIPS2016convolutional}. These spectral models usually handle the whole graph simultaneously, and are difficult to parallel or scale to large graphs. Recently, Kipf et al. designed a graph convolutional network~(GCN)  for semi-supervised learning on graph data, which can be motivated based on the spectral graph convolutional networks~\cite{IDEB2017representation,ICLR2017semi,ICLR2017graph,NIPS2017inductive}.  The key idea of GCNs is to generate node embeddings in a message passing or information diffusion manner of a graph. These GCN based models advanced previous spectral based models with much less computational cost, and could be applied to real-world  graphs.

Researchers also exploited the possibility of applying spectral models and GCNs to recommender systems. As in the collaborative setting, the user-item interaction could be defined as a bipartite graph, some works adopted the spectral graph theory for recommendation. These spectral models could leverage the overall graph structure~\cite{zheng2018spectral,monti2017geometric}. Nevertheless, these models showed high computational cost  and it is non-trivial to incorporate user~(item) features in the modeling process. As GCNs showed improved efficiency and effectiveness over the spectral models~\cite{ICLR2017semi}, a few research works exploited GCNs for recommendation~\cite{ICLR2017graph,KDD2018graph,kdir18}. These models all share the commonality of applying the graph convolution operation that aggregates the information of the graph's first-order connections.
GC-MC is one of the first few attempts that directly applied the graph convolutions  on the user-item rating graph~\cite{ICLR2017graph}. However, the graph convolution is only operated with one layer of the observed links between users and items, neglecting the higher order structure of the graph. GCMC is proposed for bipartite edge prediction with inputs of user-item interaction matrix~\cite{kdir18}. This model is consisted of two steps:  constructing  a user-user graph and item-item graph from the user-item interaction matrix, then updating user and item vectors are based on the convolutional operations of the constructed graphs. Hence, the performance of GCMC relies heavily on the user-user and item-item construction process, and the two step process is not flexible compared to the end-to-end training process. By constructing an item correlation graph, researchers developed a data-efficient GCN algorithm PinSage, which combines efficient random walks and graph convolutions to generate embeddings of nodes that incorporate both graph structure as well as node feature information~\cite{KDD2018graph}. Our work differs from them as we leverage the graph convolution operation for the recursive social diffusion in the social networks, which is quite natural. Besides, our proposed model is general and could be applied when the user~(item) attributes or the social network structure is not available.

\section{Conclusions}
In this paper, we proposed a DiffNet neural model for social recommendation. Our main contribution lies in designing
a layer-wise influence diffusion part to model how users' latent preferences are recursively influenced by the her trusted users. We showed that the proposed DiffNet model is time and storage efficient. It is also flexible when the user and item attributes are not available. The experimental results clearly showed the flexibility and effectiveness of our proposed models. E.g., DiffNet improves more than 15\% over the best baseline with NDCG@10 on \textit{Flickr} dataset. In the future, we would like to extend our model for temporal social recommendation, where the temporal changes of users' interests are implicitly reflected from their temporal feedback patterns.

\bibliographystyle{ACM-Reference-Format}

\bibliography{diffnet}

\end{document}